\documentclass[12pt]{iopart}

\usepackage{iopams}  
\usepackage{graphicx}  
\begin{document}

\title[Soft physics capabilities of CMS]
      {Soft physics capabilities of CMS in p-p and Pb-Pb}

\author{Ferenc Sikl\'er for the CMS Collaboration}

\address{KFKI Research Insitute for Particle and Nuclear Physics, Budapest, Hungary}
\ead{sikler@rmki.kfki.hu}
\begin{abstract}
The CMS experiment will provide good quality measurements of yields
and spectra of identified charged and neutral particles, both in p-p
and heavy-ion collisions, thus contributing to the study of soft
hadronic physics at the LHC energies.
\end{abstract}



The CMS experiment at the LHC is a general purpose detector designed to explore
physics at the TeV energy scale \cite{D'Enterria:2007xr}. It has a large
acceptance and hermetic coverage. The various subdetectors are: a silicon
tracker with pixels and strips ($|\eta|<2.4$); electromagnetic ($|\eta|<3$) and
hadronic ($|\eta|<5$) calorimeters; and muon chambers ($|\eta|<2.4$). The
acceptance is further extended with forward detectors: CASTOR
($5.3<|\eta|<6.6$) and Zero Degree Calorimeters ($|\eta|>8.3$). CMS detects
leptons and both charged and neutral hadrons. In the following the soft physics
capabilities are described. For an extensive review see
Ref.~\cite{D'Enterria:2007xr}.


One of the first physics results from the LHC will be the measurement of
charged hadron spectra in p-p collisions. The measurement of these basic
observables will also serve as an important tool for the calibration and
understanding of the CMS detector and will help establishing a solid basis for
exclusive physics. This example analysis uses 2 million inelastic p-p
collisions. They have been generated by the {\sc pythia} event generator
\cite{Sjostrand:2006za}.


The p-p minimum bias trigger will be based on counting towers with energy above
the detector noise level, in both forward hadronic calorimeters (HF,
$3<|\eta|<5$). A minimal number of hits (1, 2 or 3) will be required on one or
on both sides, an energy threshold value of 1.4~GeV is used in the hit
definition. Once the luminosity is high enough, events can also be taken with
the so called zero-bias trigger: a random (clock) trigger.

The Pb-Pb trigger will be similar since there are many particles produced in
the region of the forward calorimeters. In that case the event centrality will
be determined using both HF and CASTOR calorimeters in combination with the
energy measurement of forward spectator neutrons in both ZDCs.


A good measurement of differential and integrated yields requires particle
tracking down to as low $p_T$ values as possible. With a modified algorithm the
pixel detector can be employed for the reconstruction of very low $p_T$ charged
particles.  The acceptance of the method extends down to 0.1, 0.2 and
0.3~GeV/$c$ in $p_T$ for pions, kaons and protons, respectively. The obtained
proto-tracks are used for finding and fitting the primary vertex or vertices.
The measured shape and width of hit clusters are compared to the dimensions
predicted from the local direction of the trajectory. This filter helps to
eliminate incompatible trajectory candidates at an early stage.  The seeds the
trajectory building starts from are very clean, hence one seed is expected to
produce only one global track. At the end, the tracks are refitted with the
primary vertex constraint. For details see Refs.~\cite{D'Enterria:2007xr,
Sikler:2007uh, Sikler:2007sd}.

\begin{figure}
 \begin{center}
  \begin{minipage}[c]{0.52\textwidth}
  \includegraphics[width=\textwidth]{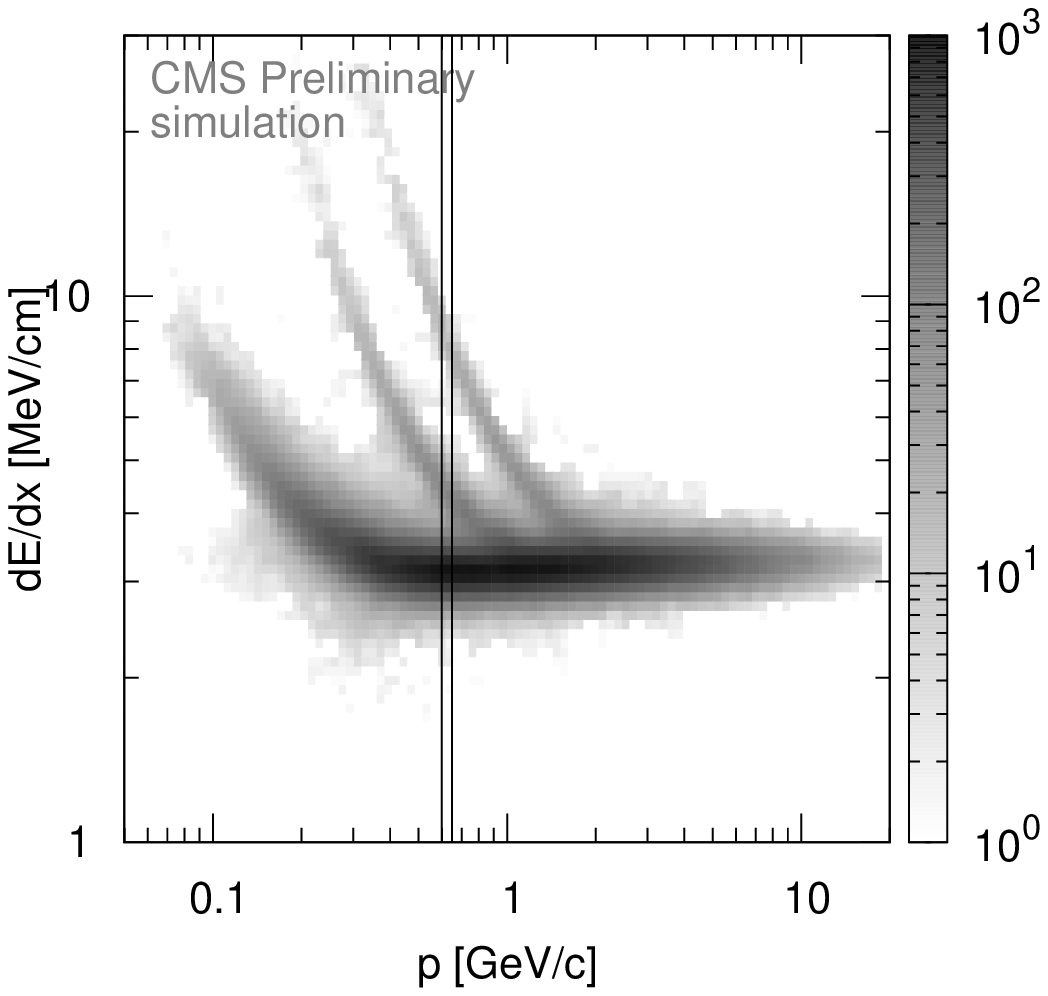}
  \end{minipage}
  \vspace{-0.05\textwidth}
  \begin{minipage}[c]{0.47\textwidth}
   \includegraphics[width=\textwidth]{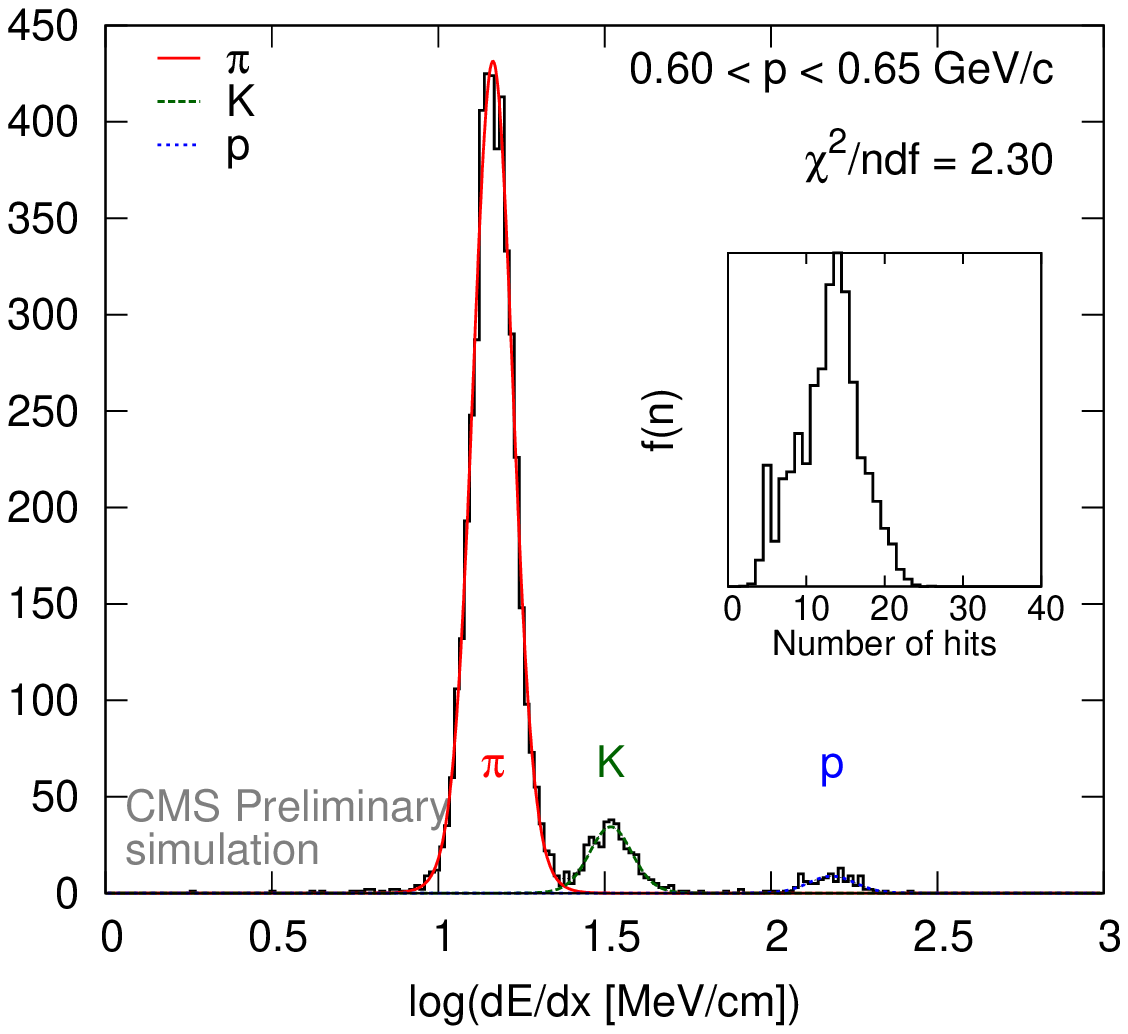}
  \end{minipage}
 \end{center}

 \caption{Left: Distribution of the truncated mean estimator
$\mathrm{d}E/\mathrm{d}x$ as a function of momentum $p$. Right: Example fit of
the $\log(\mathrm{d}E/\mathrm{d}x)$ histogram in the indicated momentum bin. The
distribution of the number of hits on track $f(n)$ is also given.}

 \label{fig:energyLoss}
\end{figure}

The hadron spectra are corrected for
feed-down from weakly decaying resonances ($\mathrm{K^0_S}$,
$\Lambda$ and $\overline{\Lambda}$).
A recorded event is used in the analysis if it passes offline track or vertex
triggers. In the case of single inelastic events, the probabilities to
reconstruct zero, one or two interaction vertices are 22\%, 74\% and 4\%,
respectively.

\begin{figure}
 \begin{center}
   \includegraphics[width=0.32\textwidth]{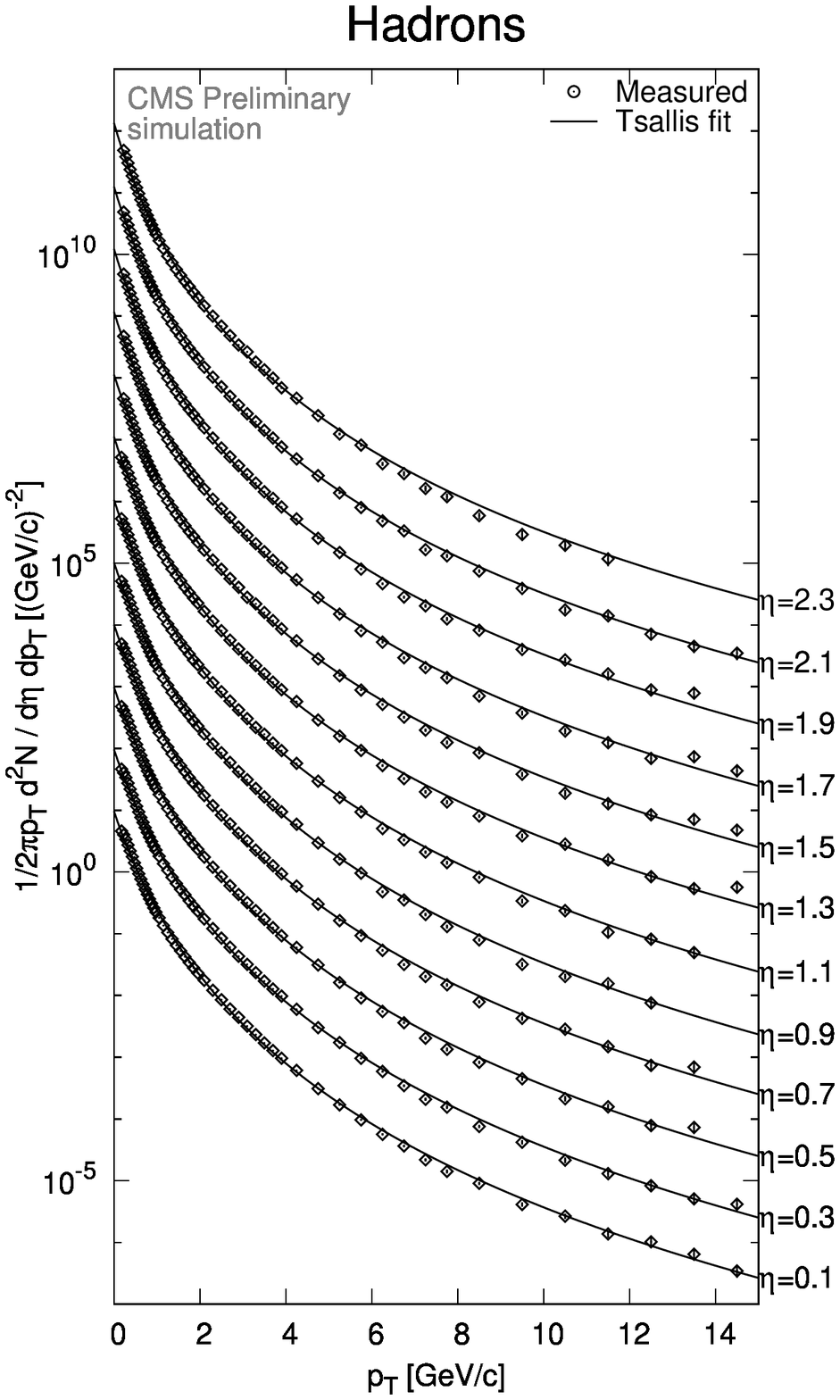}
   \includegraphics[width=0.32\textwidth]{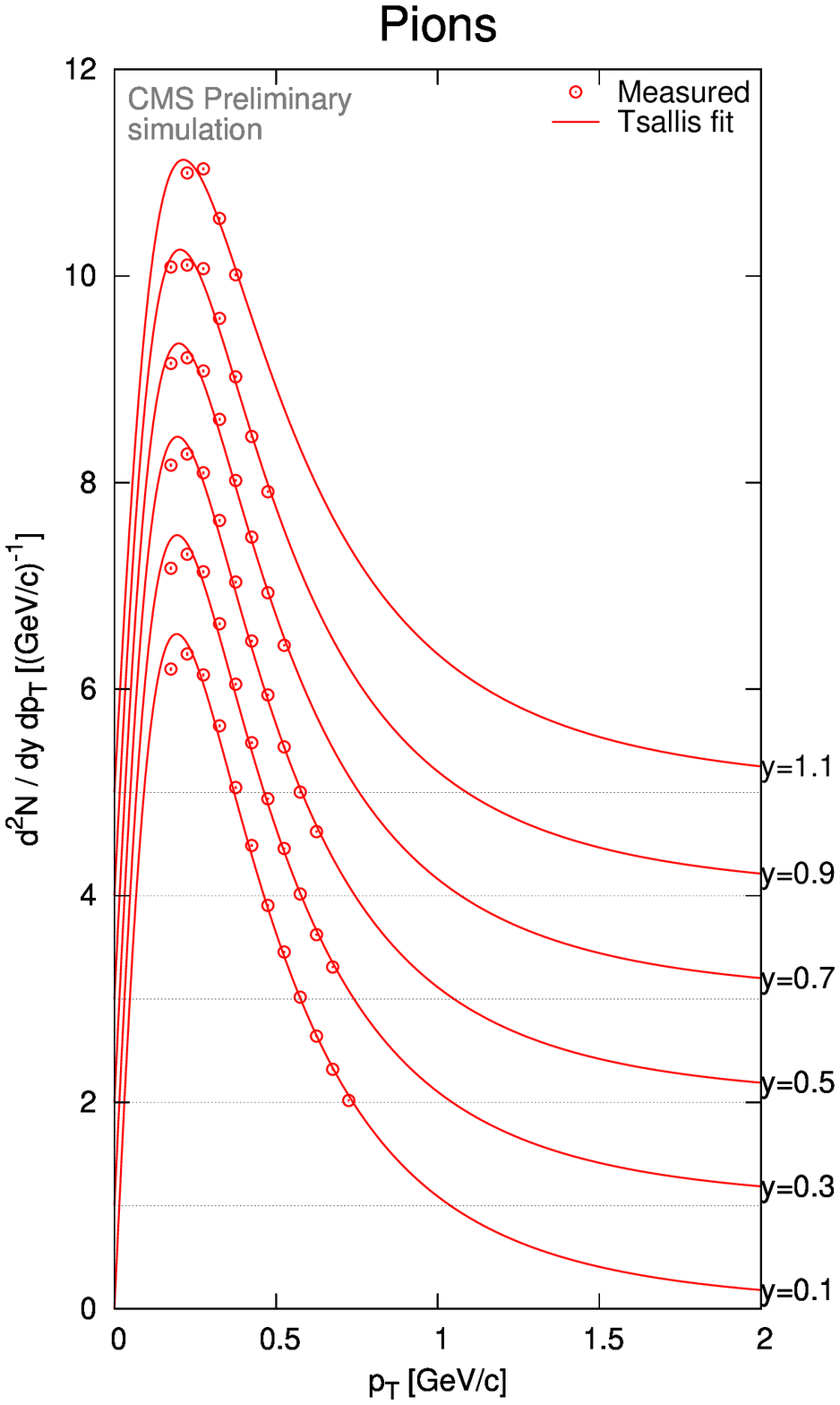}
   \includegraphics[width=0.32\textwidth]{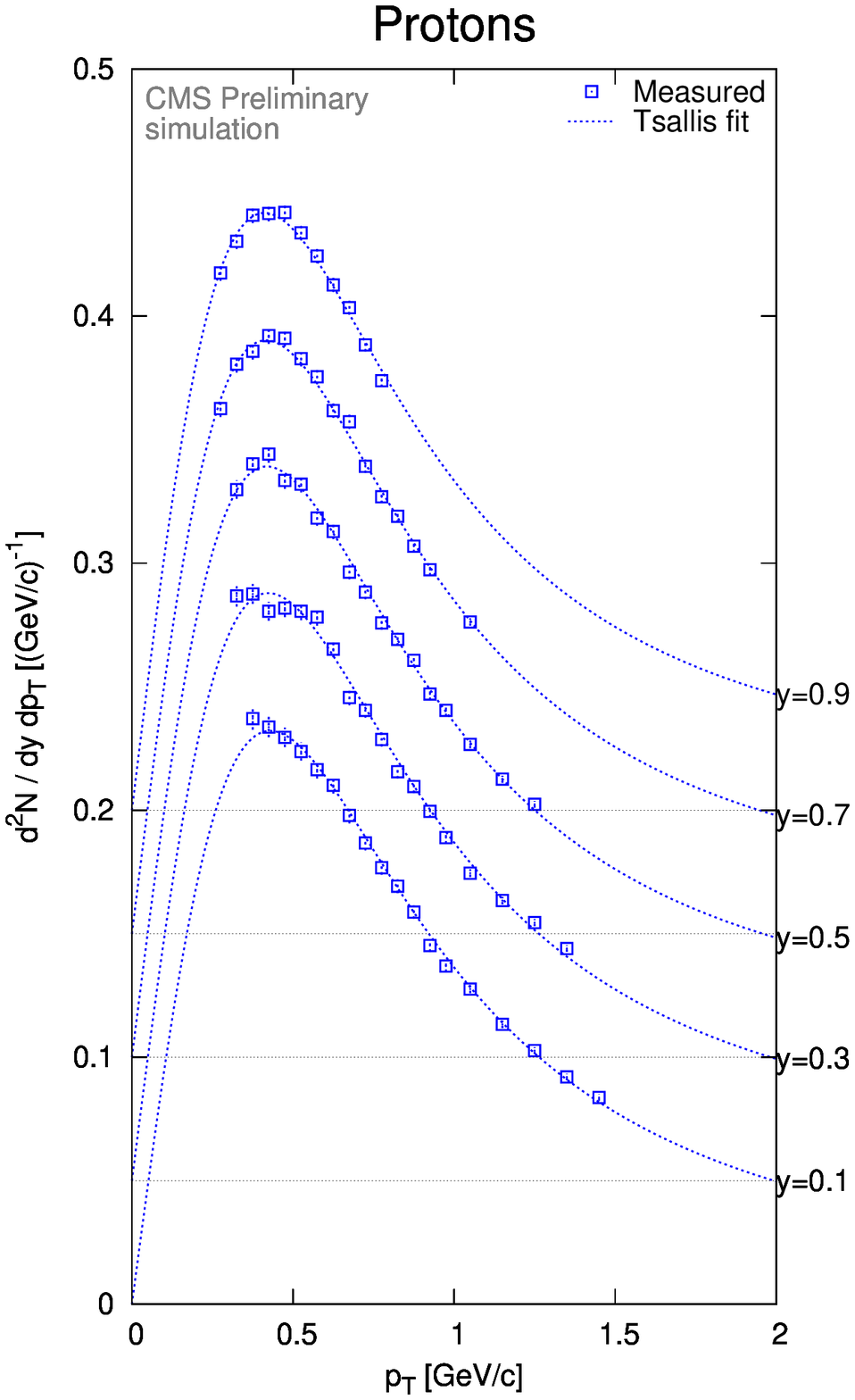}
 \end{center}

 \caption{Measured invariant and differential yields of charged hadrons, pions
and protons in a series of $0.2$ unit wide $\eta$ or $y$ bins. For charged
hadrons values are successively multiplied by 10. For identified hadrons values
are successively shifted upwards by one unit for clarity. Measured values and
empirical fit functions are plotted.}

 \label{fig:hadronSpectra}
\end{figure}


Charged particles can be singly identified or their yields can be extracted
using deposited energy in the pixel and strip silicon tracker
(Fig.~\ref{fig:energyLoss}-left). The bands of pions, kaons and protons are
well visible. The distribution of the logarithm of the estimator can be fitted
in slices of momentum (see Fig.~\ref{fig:energyLoss}-right). The relative
resolution of $\mathrm{d}E/\mathrm{d}x$ for tracks with average number of hits
($\sim$15) is around 5-7\%. The result shows that the yield of kaons can be
extracted if $p < 0.8~\mathrm{GeV}/c$ and that of protons if $p <
1.5~\mathrm{GeV}/c$. Both limits correspond to approximately $3\sigma$
separation. Details can be found in Ref.~\cite{spectra}.

\begin{figure}
 \begin{center}
  \includegraphics[width=0.49\textwidth]{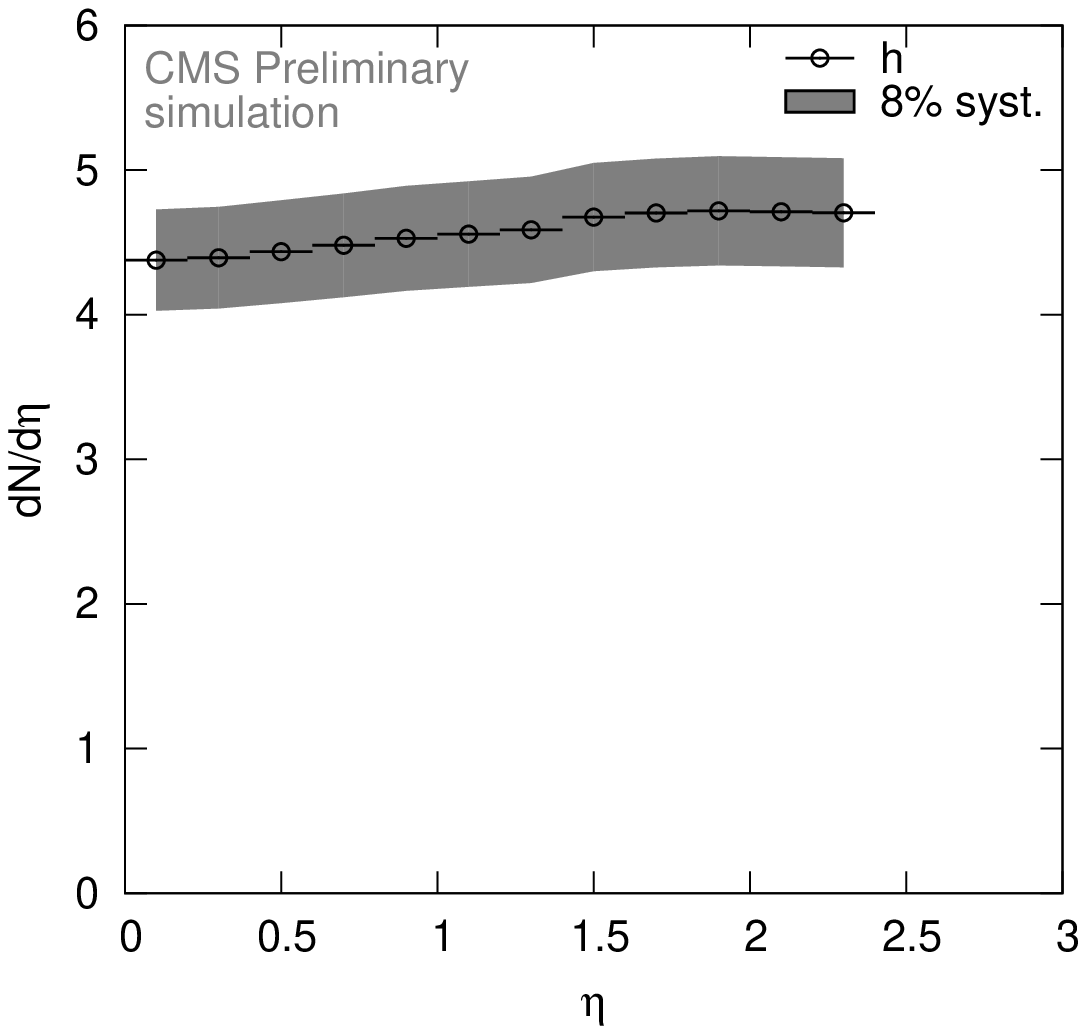}
  \includegraphics[width=0.49\textwidth]{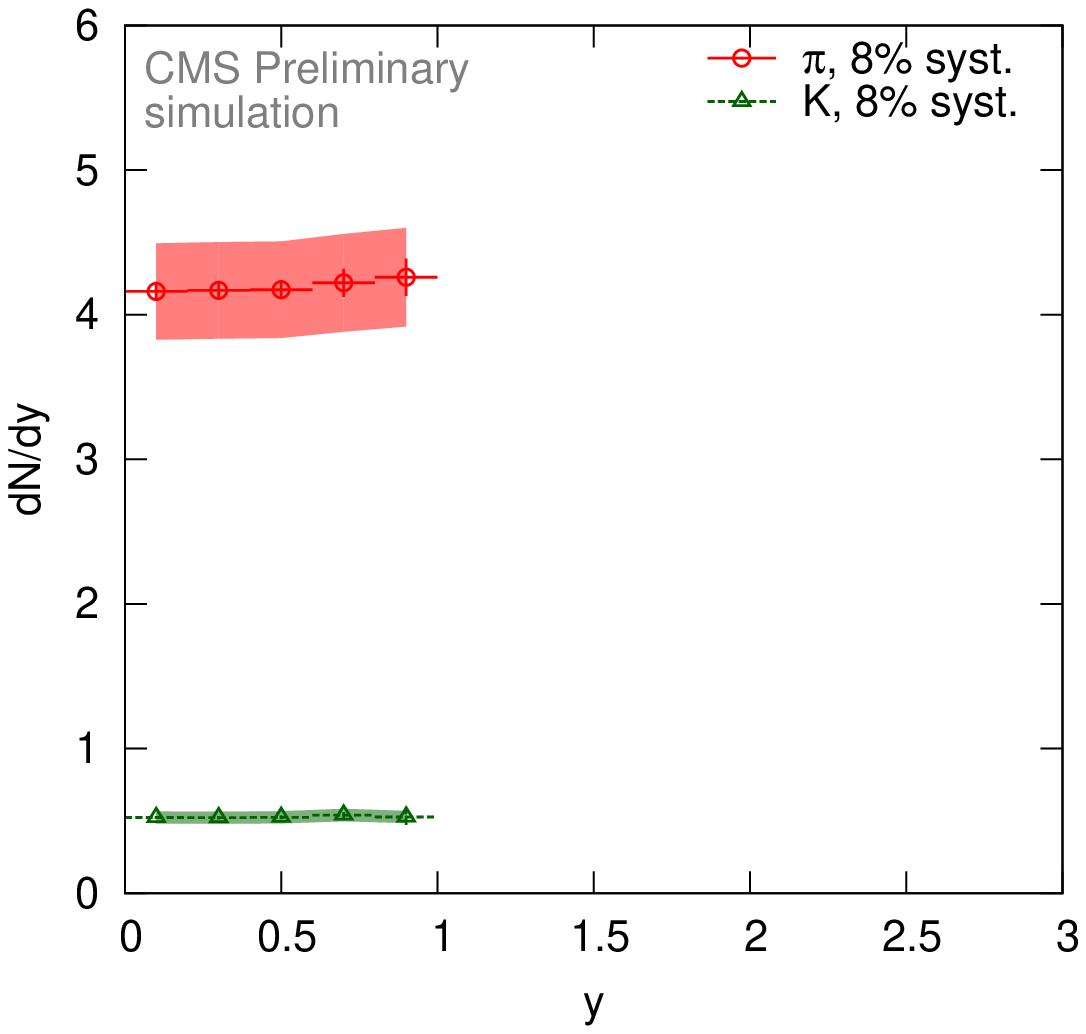}
 \end{center}

 \caption{Left: Pseudorapidity distribution of charged hadrons.  Right:
Rapidity distribution of charged pions and kaons. Estimated systematic error
bands (8\%) are also shown.}

 \label{fig:rapidityDensity}
\end{figure}
\begin{figure}
 \begin{center}
 \includegraphics[width=0.47\textwidth]{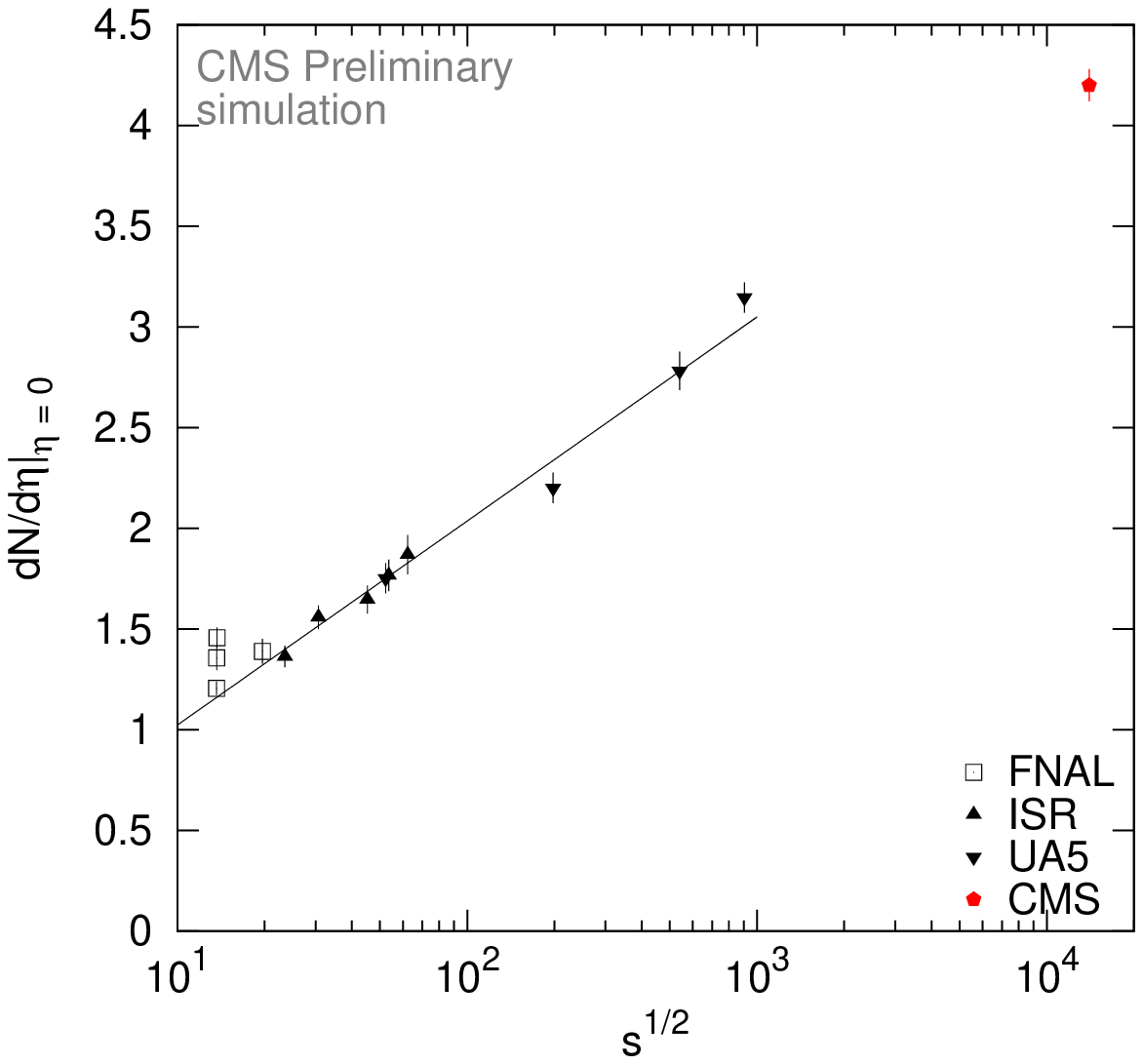}
 \includegraphics[width=0.47\textwidth]{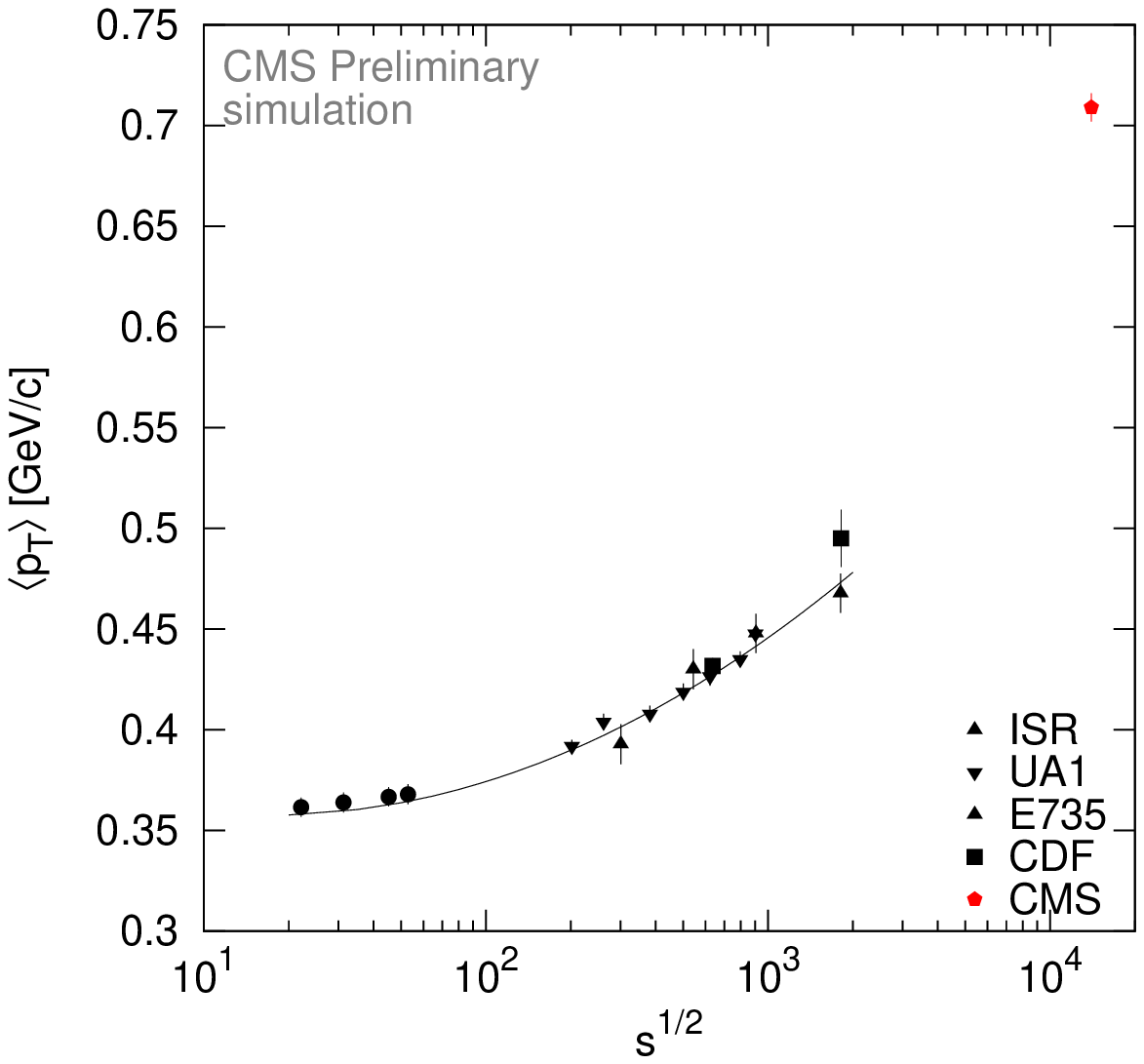}
 \end{center}

 \caption{Left: Energy dependence of pseudorapidity density of charged hadrons
at $\eta \approx 0$. Right: Energy dependence of average transverse momentum of
charged hadrons at $\eta \approx 0$. The lines show the fit to data points of
other experiments taken from Refs.~\cite{Alner:1986xu,Ullrich:2002tq}. Results
of this analysis are shown with pentagons in the top right corner.}

 \label{fig:energyDependence}
\end{figure}

The invariant yields were fitted by the Tsallis function \cite{Tsallis:1987eu}.
In general, results refer to the sum of positively and negatively charged
particles. Measured invariant and differential yields of charged hadrons are
shown in Fig.~\ref{fig:hadronSpectra}. The pseudorapidity distribution of
charged hadrons (Fig.~\ref{fig:rapidityDensity}-left) and the rapidity
distributions of pions and kaons are shown in
Fig.~\ref{fig:rapidityDensity}-right. The energy dependence of some measured
quantities can also be studied (Fig.~\ref{fig:energyDependence}). More details
can be found in Ref.~\cite{spectra}.

The reaction plane in Pb-Pb collisions can be reconstructed using
electromagnetic and hadronic calorimeters, by extracting harmonic angular
coefficients of the energy deposition distribution \cite{D'Enterria:2007xr}.
The second harmonic coefficient $v_2$ can also be determined using the silicon
tracker with an estimated systematic error below 3\%.


In summary, the CMS detector has a good capability for global event
characterization and physics with soft probes. The performance in low bias
triggering, measurement of charged hadron spectra and yields, particle
identification and elliptic flow have been shown.

\section*{Acknowledgment}

The author wishes to thank the Hungarian Scientific Research Fund and the
National Office for Research and Technology (K 48898, H07-B 74296).

\end{document}